\documentstyle[preprint,aps]{revtex}

\begin{document}
\begin{titlepage}
\rightline{CMU-HEP94-24}
\rightline{DOE-ER/40682-78}
\rightline{July, 1994}
\vspace{3cm}
\begin{center}
{\LARGE\bf CP VIOLATION IN THE DECAY $b\rightarrow s\gamma$ \\
IN THE TWO-HIGGS DOUBLET MODEL}
\end{center}
\bigskip
\begin{center}
{\large\bf  L. WOLFENSTEIN and Y. L.  WU } \\
 Department of Physics, \\ Carnegie-Mellon University, \\
Pittsburgh, Pennsylvania 15213, U.S.A.
\end{center}
\vspace{4cm}
\begin{center}
To appear in  Phys. Rev. Lett., 1994.
\end{center}

\end{titlepage}
\draft
\preprint{CMU-HEP94-24, DOE-ER/40682-78}
\title{CP Violation in the Decay $b\rightarrow s\gamma$ \\
in the Two-Higgs Doublet Model}
\author{ Lincoln Wolfenstein and Yue-Liang  Wu\cite{byline} }
\address{ Department of Physics, \ Carnegie Mellon University \\ Pittsburgh,
 Pennsylvania 15213,\ U.S.A.}
\date{July 1994}
\maketitle

\begin{abstract}
 In the most general two-Higgs doublet model with an approximate
family symmetry and
CP violation being originated solely from the relative phase of two
vacuum expectation values, CP asymmetry in the decay $b\rightarrow s\gamma$
may arise from the CP violation of the charged Higgs boson interactions
with fermions. This asymmetry may be larger than in the standard model and
can lie between $10^{-2}$ and $10^{-1}$. The decay rate of
$b\rightarrow s\gamma$ is allowed to be smaller or larger than in the
standard model.
\end{abstract}

\narrowtext

   The decay of the form $b\rightarrow s\gamma$ represents the first
observation\cite{CLEO} of a process in B decay clearly involving a loop in the
standard model. The rate of the decay is roughly that expected\cite{GSW} from
a loop involving a t-quark and $W^{\pm}$. In the simplest extension of the
standard model, the two-Higgs doublet model (2HDM), there may be a significant
contribution from the loop in which $W^{\pm}$ is replaced by the charged Higgs
boson $H^{\pm}$. Here we discuss $b\rightarrow s\gamma$ in the general 2HDM
with an emphasis on possible CP-violating efects.

     The Yukawa interaction in general in the 2HDM is

\begin{equation}
L_{Y} = \bar{q}_{L} (\Gamma_{1}^{D} \phi_{1} + \Gamma_{2}^{D}\phi_{2}) D_{R}
+ \bar{q}_{L} (\Gamma_{1}^{U} \tilde{\phi}_{1} +
\Gamma_{2}^{U}\tilde{\phi}_{2}) U_{R}
\end{equation}
where $q_{L}$ is the quark doublet and $D_{R}$ and $U_{R}$ are the right-handed
quark singlets. $\Gamma_{1}^{F}$ and $\Gamma_{2}^{F}$ are matrices in flavor
space.  The presence of both $\Gamma_{1}^{F}$ and $\Gamma_{2}^{F}$ in general
lead to flavor-changing neutral Higgs exchange (FCNE) processes. To avoid these
it is customary to choose either

\[ \mbox{Model 1}\ :  \qquad  \Gamma_{1}^{U} = \Gamma_{1}^{D} = 0 \ ; \qquad
\mbox{Model 2}\ :  \qquad  \Gamma_{1}^{U} = 0, \qquad \Gamma_{2}^{D} = 0  \]
It has been pointed out by Cheng and Sher \cite{CS} and others and reemphasized
by Hall and Weinberg \cite{HW} that FCNE may be suppressed simply by an
approximate flavor symmetry allowing both $\Gamma_{1}^{F}$ and $\Gamma_{2}^{F}$
to be significant. The consequences of an assumption of approximate global U(1)
family symmetries (AGUFS) (i.e., one for each family) have been
worked out in detail in \cite{YLWU} and emphasized recently in \cite{WW}.
AGUFS are sufficient for a natural suppression of family-changing currents
(for both charged and neutral currents). In particular as we have pointed out
\cite{YLWU,WW} this has important consequences for the charged Higgs
couplings allowing significant new CP-, T- and P-violating effects on both
indirect CP violation ($\epsilon$) and direct CP violation
($\epsilon'/\epsilon$) in kaon decay and electric dipole
moments of the neutron ($D_{n}$) and electron ($D_{e}$).

   Our other  assumption is that all significant CP violation arises
from  the Higgs vacuum expectation values

\begin{equation}
< \phi_{1} > = \frac{v}{\sqrt{2}} \cos\beta e^{i\delta} \ ,  \qquad
< \phi_{2} > = \frac{v}{\sqrt{2}} \sin\beta
\end{equation}
This leads to a phase in the CKM matrix as in the standard model but provides
a varity of new sources of CP violation.

   The important $H^{\pm}$ couplings can be obtained neglecting FCNE and
only considering the standard CKM quark mixing. Neglecting the small
off-diagonal terms, one finds for the top mass

\begin{equation}
 m_{t} e^{-i\delta_{t}} = (h_{1}\cos\beta e^{-i\delta} + h_{2}\sin\beta) v
\end{equation}
where $h_{1}(h_{2})$ are the 33 diagonal elements of $\Gamma_{1}^{U}$,
$\Gamma_{2}^{U}$. The phase $\delta_{t}$ must be removed by rephasing $t_{R}$.
The coupling of $H^{\pm}$ to $t_{R}$ then has the form

\begin{equation}
-H^{-} \bar{b}_{L j} V^{\dagger}_{ji} t_{R i} (h_{1}\sin\beta e^{-i\delta} -
 h_{2}\cos\beta) v e^{i\delta_{t}} \equiv - \xi_{t} m_{t}
H^{-} \bar{b}_{L j} V^{\dagger}_{ji} t_{R i}
\end{equation}
where $V$ is the CKM matrix. A similar equation holds for the $b_{R}$
couplings.
It is easy to show that the coefficients $\xi_{t}$, $\xi_{b}$ can be written

\begin{equation}
\xi_{f} = \frac{\sin\delta_{f}}{\sin\beta\cos\beta\sin\delta}
e^{i\sigma_{f} (\delta - \delta_{f})} - \cot\beta
\end{equation}
where $\sigma_{f} = +$ for $b$ and $\sigma_{f} = -$ for $t$ and $\delta_{f}$
serves to parameterize the ratio  $h_{2}/h_{1}$. In the limiting case of
Model 1, $\delta_{t} = \delta_{b} = 0$, i.e.,
$\xi_{t} = \xi_{b} = - \cot\beta$.
In the case of Model 2, $\delta_{b} = \delta$,  $\delta_{t} = 0$,
i.e., $\xi_{b} = \tan\beta $, $\xi_{t} = - \cot\beta$.
In general Eq.(5) can be seen as interpolating between the values $\tan\beta$
and $-\cot\beta$; however outside of these limits $\xi_{f}$ has a complex
phase.

   Considering the leading loop diagram,
the decay amplitude of $b\rightarrow s\gamma$ can be written

\begin{equation}
  < s \gamma | T | b > \equiv {\cal T}_{s\gamma} =
 v_{t}(A_{s\gamma}^{W} + \xi_{t}\xi_{b} A_{s\gamma}^{H}
 + \xi_{t}\xi_{t}^{\ast} \tilde{A}_{s\gamma}^{H})
\end{equation}
where $v_{t} = V_{tb}V_{ts}^{\ast}$. $A_{s\gamma}^{W}$, $A_{s\gamma}^{H}$
and  $\tilde{A}_{s\gamma}^{H}$
can be generally expressed

\begin{equation}
A_{s\gamma}^{W,H}(t) =  C_{s\gamma}^{W,H} O_{s\gamma}, \qquad
\tilde{A}_{s\gamma}^{H}(t) =  \tilde{C}_{s\gamma}^{H} O_{s\gamma}
\end{equation}
with
\begin{equation}
O_{s\gamma}  = -\frac{G}{8\sqrt{2} \pi^{2}} e \bar{u}_{s}(p) \sigma^{\mu\nu}
(1+\gamma_{5}) u_{b}(p_{b}) F_{\mu\nu}
\end{equation}
and $C_{i}$ the Wilson coefficient functions \cite{GSW}
\begin{eqnarray}
C_{s\gamma}^{W} & = & -\eta^{16/23}
[ \frac{1}{2} A(x_{t}) + \frac{4}{3} D(x_{t}) (\eta^{-2/23} -1) +
\frac{232}{513} (\eta^{-19/23}-1) ] \nonumber \\
C_{s\gamma}^{H} & = & \eta^{16/23}
[ B(y_{t}) + \frac{8}{3} E(y_{t}) (\eta^{-2/23} -1) ] \\
\tilde{C}_{s\gamma}^{H} & = & -\eta^{16/23}
\frac{1}{6} [A(y_{t}) + \frac{8}{3} D(y_{t}) (\eta^{-2/23} -1) ] \nonumber
\end{eqnarray}
where $A$, $B$, $D$ and $E$ are the
integral functions of $x_{t} =m_{t}^{2}/m_{W}^{2}$ and $y_{t} =
m_{t}^{2}/m_{H^{+}}^{2}$ and defined in Ref.\cite{GSW} and $\eta =
\alpha_{s}(m_{W})/\alpha_{s}(m_{b})$.

  With these considerations, we now evaluate the decay rate of $b\rightarrow
s\gamma$. To reduce the uncertainties from the CKM matrix elements and
b-quark mass, one usually relates $BR(b\rightarrow s\gamma)$ to the inclusive
semileptonic decay rate $\Gamma(b\rightarrow c e \nu)$ \cite{HBBP,BURAS}
\begin{eqnarray}
BR(b\rightarrow s\gamma) & = & [\frac{\Gamma (b\rightarrow s\gamma)}{
\Gamma (b\rightarrow c e\nu)}]^{th} \times BR^{exp}(b\rightarrow c e\nu)
\nonumber \\
& \simeq & 0.031 \{ [C_{s\gamma}^{WH} +
Re(\xi_{t}\xi_{b}) C_{s\gamma}^{H}]^{2} +
[Im(\xi_{t}\xi_{b}) C_{s\gamma}^{H}]^{2} \} BR^{exp}(b\rightarrow c e\nu)
\end{eqnarray}
where $C_{s\gamma}^{WH} = C_{s\gamma}^{W} + |\xi_{t}|^{2}
\tilde{C}_{s\gamma}^{H} $.

  In general $\xi_{t}$ is expected to be of order unity or less if the Yukawa
couplings of the top quark are reasonable. In this case, the term proportional
to $|\xi_{t}|^{2}$ makes only a small contribution. In our numerical examples
and figures we let $|\xi_{t}|^{2}=0$, but results with $|\xi_{t}|^{2}=1$
are very similar. On the other hand $\xi_{b}$ may have a
magnitude considerably larger, as in the limiting case of model 2 with large
$\tan\beta$. In Figure 1 we show the ratio of the 2HDM result for the branching
ratio to that for the standard model as a function of $Re (\xi_{b}\xi_{t})$
and $Im (\xi_{b}\xi_{t})$ for $m_{H^{+}} = m_{t}$. For quite resonable values
the result can be either greater or smaller than in the standard model.
This is in contrast to model 2 in which the rate is always greater than in the
standard model. The possibility of a smaller value has also been noted in
model 1 \cite{HBBP}.

  If it is established that the rate for $b\rightarrow s + \gamma$ differs
from that in the standard model, this could be the first indication of the
existence of a charged Higgs boson, at the present, however, there is
considerable uncertainty \cite{ROMA} in the standard model rate. In the
2HDM discussed here, however, there is the possibility of a more distinct
signature of the charged Higgs contribution  due to CP violation. To calculate
the CP-violating rate difference between $B$ and $\bar{B}$ decays it is
necessary to include final-state-interaction effects.
Using the general formalism in \cite{LW}, the decay amplitude of
$b\rightarrow s\gamma$ in eq.(6) is modified by including the corresponding
absorbtive terms via

\begin{eqnarray}
 {\cal T}_{s\gamma} & = &  v_{t} (A_{s\gamma}^{W} + i A_{sg}^{W}
\ t_{sg\rightarrow s\gamma}) + \sum_{q}^{u,c} v_{q} i A_{sq\bar{q}}^{W}(q)
\ t_{sq\bar{q}\rightarrow s\gamma} \nonumber \\
& & + v_{t} [ \xi_{t}\xi_{b} (A_{s\gamma}^{H} + i A_{sg}^{H}
\ t_{sg\rightarrow s\gamma}
 + \xi_{t}\xi_{t}^{\ast} (\tilde{A}_{s\gamma}^{H}
 + i \tilde{A}_{sg}^{H} \  t_{sg\rightarrow s\gamma})]
\end{eqnarray}
where $v_{q} = V_{qb}V_{qs}^{\ast}$ are products of CKM elements and
$t_{i\rightarrow s\gamma}$ is the scattering amplitudes. $A_{sg}^{W,H}$ and
$\tilde{A}_{sg}^{H}$ are expressed

\begin{equation}
A_{sg}^{W,H}  =  C_{sg}^{W,H} O_{sg}, \qquad
\tilde{A}_{sg}^{H}(t) = \tilde{C}_{sg}^{H} O_{sg}
\end{equation}
with
\begin{equation}
O_{sg}  = -\frac{G}{8\sqrt{2} \pi^{2}} g \bar{u}_{s}(p) \sigma^{\mu\nu}
T^{a} (1+\gamma_{5}) u_{b}(p_{b}) G_{\mu\nu}^{a}
\end{equation}
and $C_{i}$ the Wilson coefficient functions \cite{GSW,BURAS}
\begin{eqnarray}
C_{sg}^{W} & = & -\eta^{14/23}
[ \frac{1}{2} D(x_{t}) + 0.1687]  \nonumber \\
C_{sg}^{H} & = & \eta^{14/23}E(y_{t}), \qquad \tilde{C}_{sg}^{H}  =
-\eta^{14/23}\frac{1}{6} D(y_{t})
\end{eqnarray}
where $\eta = \alpha_{s}(m_{W})/\alpha_{s}(m_{b})\simeq 0.56$ for
$\Lambda_{QCD} = 150$ MeV and $m_{b} =4.9$ GeV.

$A_{sq\bar{q}}$ is the $b\rightarrow sq\bar{q}$ amplitude
\begin{equation}
A_{sq\bar{q}} = - \frac{G}{\sqrt{2}} c_{1} \bar{u}_{q}(p_{1})\gamma_{\mu}
(1-\gamma_{5})u_{b}(p_{b}) \bar{u}_{s}(p') \gamma^{\mu}(1-\gamma_{5})
v_{\bar{q}}(p_{2})
\end{equation}
where $c_{1}$ is the QCD correction with $c_{1}=1.1$.

The rate difference is calculated to be
\begin{eqnarray}
\Delta_{s\gamma} & \equiv & \Gamma (\bar{b}\rightarrow \bar{s}\gamma ) -
\Gamma (b\rightarrow s\gamma ) \nonumber \\
& = & 4 |v_{t}|^{2} (C_{s\gamma}^{H} C_{sg}^{WH} -
C_{s\gamma}^{WH} C_{sg}^{H} ) O_{s\gamma}^{\dagger} O_{sg}
t_{sg\rightarrow s\gamma} Im (\xi_{t}\xi_{b})   \\
& & + \sum_{q}^{u,c} 4 C_{s\gamma}^{WH} O_{s\gamma}^{\dagger} A_{sq\bar{q}}
t_{sq\bar{q}\rightarrow s\gamma} Im (v_{t}v_{q}^{\ast}) +
\sum_{q}^{u,c} 4 C_{s\gamma}^{H} O_{s\gamma}^{\dagger} A_{sq\bar{q}}
t_{sq\bar{q}\rightarrow s\gamma} Im (v_{t}v_{q}^{\ast} \xi_{t}\xi_{b})
\nonumber
\end{eqnarray}
Here we have omitted the flux factor and the phase space integrals.
As shown by Soares \cite{JS} that after integration over the phase
space variables for $\Delta_{s\gamma}$, the absorbtive term with
the $sq\bar{q}$ intermediate state is suppressed by a factor
of about $\alpha_{s}/4$ for the up-qaurk and for the charm-quark it has
an additional suppression factor of about $0.12$ from  phase space.
For the absorbtive term with the $sg$ intermediate state, there is
no extra phase space suppression and its magnitude after integration over the
phase space variables is expected to be suppressed just by
a factor of order $\alpha_{s}$.

With these considerations, we then obtain the CP asymmetry observable

\begin{eqnarray}
A^{CP}_{s\gamma} & = & \frac{\Delta_{s\gamma}}{2\Gamma(b\rightarrow s\gamma)}
\simeq  \frac{(C_{s\gamma}^{H} C_{sg}^{WH} -
C_{s\gamma}^{WH} C_{sg}^{H})\alpha_{s}}{C_{s\gamma}^{2}} Im (\xi_{t}\xi_{b})
  \nonumber \\
& & - \frac{0.12 Re(v_{t}v_{c}^{\ast}) + Re(v_{t}v_{u}^{\ast})}{|v_{t}|^{2}}
\frac{C_{s\gamma}^{H}c_{1} \alpha_{s}}{2C_{s\gamma}^{2}} Im(\xi_{t}\xi_{b}) \\
& & + \frac{Im(v_{t}v_{u}^{\ast})}{|v_{t}|^{2}}
\frac{(C_{s\gamma}^{WH}+ Re(\xi_{t}\xi_{b})C_{s\gamma}^{H})
c_{1} \alpha_{s}}{2C_{s\gamma}^{2}} \nonumber
\end{eqnarray}
where the first two terms arise from the new source of CP violation for the
charged-Higgs boson interactions of the fermions with intermediate states
$sg$ and $sq\bar{q}$ ($q=u,c$), respectively, and the last term arises from the
CKM phase. This last term has been analyzed in detail by Soares \cite{JS} in
the standard model with the resulting asymmetry between $10^{-2}$ and
$10^{-3}$.
Our major interest lies in the first two terms which can result in a much
larger asymmetry than in the standard model. Values of the asymmetry
considering these terms alone for various values of $Re(\xi_{t}\xi_{b})$ and
$Im(\xi_{t}\xi_{b})$ are illustrated in Figure 1. It is seen that asymmetries
between 2 and 10 \% are quite reasonable.  These results as those in the
standard model are necessarily uncertain because of the use of quark diagrams
to calculate the final state interaction effect. However an asymmetry well
above 1\% would be a strong indication of this new physics.

   It is important to look at effects of $\xi_{t}\xi_{b}$ on other observables.
In the case of $Re(\xi_{t}\xi_{b})$ there is no observable that is as sensitive
as the $b\rightarrow s + \gamma$ rate. On the other hand $Im (\xi_{t}\xi_{b})$
is relevant for the neutron electric dipole moment $D_{n}$ due to the
Weinberg dimension-6 gluonic operator\cite{SW}. This operator can be
induced from two loop diagrams through the charged  Higgs boson exchange
with  internal loop top- and bottom-quarks and three external
gluons \cite{DICUS}. In the Weinberg three-Higgs doublet model,
the CP-violating phase arises from the mixings among the charged Higgs
bosons,  which is not relevant in the 2HDM. In the version of the 2HDM
discussed here the CP-violating phase comes from the
complex Yukawa couplings $\xi_{t}$ and $\xi_{b}$ giving

\begin{equation}
d_{n} = (3.3-0.11) \times 10^{-25} Im(\xi_{t}\xi_{b}) [12 h_{C}(y_{b},y_{t})]
\mbox{e cm}
\end{equation}
where the first value in the bracket is from the naive dimensional analysis
\cite{SW} and the second value in the bracket is from a recent reanalysis
\cite{BU} for the hadronic matrix element. $h_{C}$ is an integral function
\cite{DICUS} with $h_{C} = 1/12$ for $m_{H^{+}}= m_{t}$. From the present
experimental upper limit for the neutron electric dipole moment
$d_{n} < 1.2 \times 10^{-25}$ e cm, it is seen that the limit on
$Im(\xi_{t}\xi_{b})$ ranges from $0.3-10$ due to the large uncertainties
of the hadronic matrix element. This limit still allows the large asymmetries
discussed above. On the other hand the detection of such a CP-violating
asymmetry in $b\rightarrow s \gamma$ would indicate that $d_{n}$ is not too
far below the present limit.

  In conclusion, we have shown that in the most general two-Higgs doublet
model\cite{YLWU,WW}  with approximate global U(1) family symmetries (AGUFS)
and CP violation being originated solely from a single relative phase of two
vacuum expectation values, the CP asymmetry in the decay $b\rightarrow s\gamma$
due to new sources of CP violation for charged-Higgs boson interactions and
final-state-interaction effects can lie between $10^{-2}$ and $10^{-1}$
which is one order of magnitude larger than in the standard model.

This work was supported by DOE  grant \# DE-FG02-91ER40682.

\vspace{15cm}

\begin{figure}
\caption{ The solid lines show the ratio $r =\Gamma_{2HDM}(b\rightarrow s
\gamma)/\Gamma_{SM}(b\rightarrow s \gamma)$ as a function of
$Im(\xi_{t}\xi_{b})$ and $Re(\xi_{t}\xi_{b})$.
The three solid curves are for $r=0.5, 1, 1.5$ which correspond to
radii $R = 1.56, 2.2, 2.69$, respectively, with the central point at
$Im(\xi_{t}\xi_{b})=0$ and $Re(\xi_{t}\xi_{b})=2.2$  and
$m_{H^{+}} = m_{t}$.  The dotted lines are for CP asymmetry $A_{s\gamma}^{CP}=
2\% $, dashed lines are for $A_{s\gamma}^{CP}= 5\%$ and dot-dashed line is
for $A_{s\gamma}^{CP}= 10\%$ with $m_{H^{+}} = m_{t}$.}
\label{autonum}
\end{figure}


\begin{references}
\bibitem[*]{byline}  On leave of absence from University Dortmund, Germany. \\
After October 15, 1994, Department of Physics, Ohio State University.
\bibitem{CLEO} R. Ammar et. al. (CLEO Collaboration), Phys. Rev.
Lett. {\bf 71}, 674 (1993).
\bibitem{GSW} B. Grinstein, R. Springer and M.B. Wise, Nucl. Phys. {\bf B339},
269 (1990); A. Ali and C. Greub, Phys. Lett. {\bf B259}, 182 (1991).
\bibitem{CS} T.P. Cheng and M. Sher, Phys. Rev. {\bf D35}, 3484 (1987).
\bibitem{HW} L.J. Hall and S. Weinberg, Phys. Rev. {\bf D48}, 979 (1993).
\bibitem{YLWU} Y.L. Wu, `` A Model of CP Violation", pp 80, CMU-HEP94-01,
hep-ph/9404241, 1994; `` Origin and Mechanisms of CP Violation",
CMU-HEP94-02, hep-ph/9404271, 1994; see also ``A Model for the Origin and
Mechanisms of CP Violation", invited talk at the 5th Conference on the
Intersections of Particle and Nuclear Physics, May 31-June 6, 1994, at St.
Petersburg, Florida (to appear in the Proceedings), CMU-HEP94-17,
hep-ph/9406306.
\bibitem{WW} Y.L. Wu and L. Wolfenstein, ``Sources of CP Violation in the
Two-Higgs Doublet Model", Phys. Rev. Lett. {\bf 73},
1762 (1994).
\bibitem{HBBP} J.L. Hewett, Phys. Rev. Lett. {\bf 70}, 1045 (1993);
V. Barger, M.S. Berger and R.J.N. Phillips, Phys. Rev. Lett. {\bf 70},
1368 (1993).
\bibitem{BURAS} A.J. Buras, M. Misiak, M. M\"{u}nz and S. Pokorski,
MPI-Ph/93-77, TUM-T31-50/93, 1993.
\bibitem{ROMA} M. Ciuchini, E. Franco, G. Martinelli, L. Reina and
L. Silvestrini, CERN-TH.7283/94, ROME prep. 94/1020, ULB-TH 09/94,
hep-ph/9406239, 1994.
\bibitem{LW} L. Wolfenstein, Phys. Rev. {\bf 43D}, 151 (1991).
\bibitem{JS} J.M. Soares, Nucl. Phys. {\bf B367}, 575 (1991).
\bibitem{SW} S. Weinberg, Phys. Rev. Lett. {\bf 63}, 2333 (1989); Phys. Rev.
{\bf D42}, 860 (1990).
\bibitem{DICUS} D.A. Dicus, Phys. Rev. {\bf D41}, 999 (1990).
\bibitem{BU} I.I. Bigi and N.G. Uraltsev, Nucl. Phys. {\bf B353}, 321 (1991).
\end{references}
\end{document}